\documentclass[fleqn,10pt]{wlscirep}
\usepackage[utf8]{inputenc}
\usepackage[T1]{fontenc}
\usepackage{bm}
\usepackage{braket}
\usepackage{multirow}
\title{Semiconductor Qubits In Practice}

\newcommand {\sivm} {SiV$^-$}
\newcommand {\nvm} {NV$^-$}
\newcommand {\sivo} {SiV$^0$}
\newcommand{\erb}{Er$^{3+}$}
\DeclareUnicodeCharacter{2212}{-}

\author[1,*]{Anasua Chatterjee}
\author[2]{Paul Stevenson}
\author[3]{Silvano De Franceschi}
\author[4]{Andrea Morello}
\author[2]{Nathalie de Leon}
\author[1,*]{Ferdinand~Kuemmeth}
\affil[1]{Center for Quantum Devices, Niels Bohr Institute, University of Copenhagen, 2100 Copenhagen, Denmark}
\affil[2]{Department of Electrical Engineering, Princeton University, Princeton, NJ 08544, USA}
\affil[3]{Universit\'e Grenoble Alpes and CEA, IRIG/PHELIQS, F-38000 Grenoble, France}
\affil[4]{Centre for Quantum Computation \& Communication Technology, School of Electrical Engineering \& Telecommunications, UNSW Sydney, NSW2052, Australia}
\affil[*]{e-mail: kuemmeth@nbi.dk}

\begin{abstract}
In recent years semiconducting qubits have undergone a remarkable evolution, making great strides in overcoming decoherence as well as in prospects for scalability, and have become one of the leading contenders for the development of large-scale quantum circuits. In this Review we describe the current state of the art in semiconductor charge and spin qubits based on gate-controlled semiconductor quantum dots, shallow dopants, and color centers in wide band gap materials. We frame the relative strengths of the different semiconductor qubit implementations in the context of quantum simulations, computing, sensing and networks. By highlighting the status and future perspectives of the basic types of semiconductor qubits, this Review aims to serve as a technical introduction for non-specialists as well as a forward-looking reference for scientists intending to work in this field.
\end{abstract}
\begin{document}

\flushbottom
\maketitle

\thispagestyle{empty}

\noindent \textbf{Key points:} 
\begin{itemize}
\item Semiconductor qubits span an entire ecosystem and are extremely versatile in terms of quantum applications, particularly viewed through the lenses of quantum simulation, sensing, computation, and communication.
\item Controlling the charge degree of freedom in gated quantum dots is important for sensing of quantum objects, readout and light-matter coupling.
\item Gate-controlled spin qubits have demonstrated long coherence times, fast two-qubit gates, and fault-tolerant operation, with promising prospects for quantum computation.
\item Shallow dopants have shown the longest coherence times in the solid state, as well as high sensitivity to magnetic fields, relevant for quantum memories and sensing.
\item Optically active defects have shown great promise as in-situ sensors, and their natural ability to serve as spin-photon interfaces makes them suitable to long-distance quantum communication.
\item Looking beyond a fault-tolerant quantum computer, semiconductor qubits will find diverse applications such as light-matter networks, scanning sensors, quantum memories, global cryptographic networks, and small-scale designer simulation arrays.
\end{itemize}

\noindent \textbf{Website summary:} Future quantum applications for semiconductor qubits will be diverse, depending strongly on the specific properties of the underlying quantum hardware. This Review considers semiconductor qubit implementations from the perspective of an ecosystem of quantum-enabled applications, such as simulation, sensing, computation, and communication.

\section*{Introduction}

The world of quantum devices has changed at a breakneck pace over the last few decades. In particular, the field has witnessed the development of a large variety of quantum bits or qubits, i.e. quantum two-level systems whose state can be initialized, coherently controlled, and measured with high fidelity. Increasing research efforts are focusing on physical implementations and materials offering the most serious prospects for large-scale integration. Correspondingly, scaling and engineering efforts are increasingly moving from academic research labs to industrial-scale development centers. Semiconductor materials have been at the forefront of this revolution, along with other quantum technologies based on ion traps and superconducting circuits.

Quantum applications will be far reaching and will depend strongly on the specific properties of the underlying quantum hardware. 
Therefore, the focus of this review goes beyond quantifying the merits of different semiconductor qubits as per their potential role in universal quantum computers. We also explore their relevance for different applications that make use of their quantum character, including sensing, simulation, computation and communication.

The field of semiconductor qubits itself spans a variety of systems, material implementations and techniques. The semiconductor qubits demonstrated so far differ from each other in various ways; they vary from systems that operate at millikelvin temperatures, achievable only inside dilution refrigerators, to systems that are suitable for room-temperature operation. They can be artificially engineered potential wells confining quantized electronic states, or single-atom impurities in a lattice, exploiting either nuclear or electronic degrees of freedom. Despite these differences, however, they share certain properties, such as their potential for high-density integration on a large scale, which originates from the well-established nanofabrication technology of the semiconductor industry. Some semiconductor qubits also boast some of the longest coherence times ever reported. Following material and technology developments in the past few years, silicon spin qubits were able to meet all of the DiVincenzo criteria, and paradigmatic two-qubit quantum algorithms have been demonstrated.  Color centers have demonstrated long-range entanglement generation capabilities and have shown themselves to be sensitive probes for nanoscale magnetometry at room temperature, with particular relevance for biomedicine.

This Review reflects the wide variety of platforms offered by different semiconductor qubit systems. Because of the extremely rapid growth of this field in the past few years, here we summarize the state of art of the field, rather then its historic development, with emphasis on relating different categories of semiconducting qubit implementations to their respective strengths and prospects for practical applications. Specifically, we will describe semiconductor qubits based on charge and spin degrees of freedom. Among spin-based qubits, we shall devote the largest attention to gate-controlled quantum dots, dopants and color centers, with the qubit encoded in an individual electronic or nuclear spin. Other qubit realizations exist where quantum information is encoded in a multi-electron state, such as the singlet-triplet qubit, the hybrid spin-charge qubit, or the flip-flop spin qubit. We shall not discuss the case of semiconductor self-assembled quantum dots on which extensive literature can be found (these systems are particularly relevant for the realization of single-photon sources\cite{Michler2017}). In the interest of concision, we also exclude from this review gatemon qubits and topological qubits, even though semiconductors are at the heart of their operating principles. We refer to recent reviews for superconducting qubits\cite{Kjaergaard2020} and Majorana zero modes\cite{Lutchyn2018}, respectively.

Each qubit category is benchmarked against four quantum technology applications. In quantum sensing applications, some observable of the qubits is sensitive to the desired environmental variable, without perturbing it in a manner that cannot be corrected for. In quantum simulations, the Hamiltonian of a physical system of interest is mapped onto the Hamiltonian of appropriately controlled qubit circuit. For quantum computing applications, we assess the fidelity, prospects for two-qubit control, and coherence time of the qubit. For quantum communication, semiconductor qubits can serve as nodes in quantum networks, enabling non-classical communication between distant sites.

\begin{figure*}
	\includegraphics[width=\textwidth]{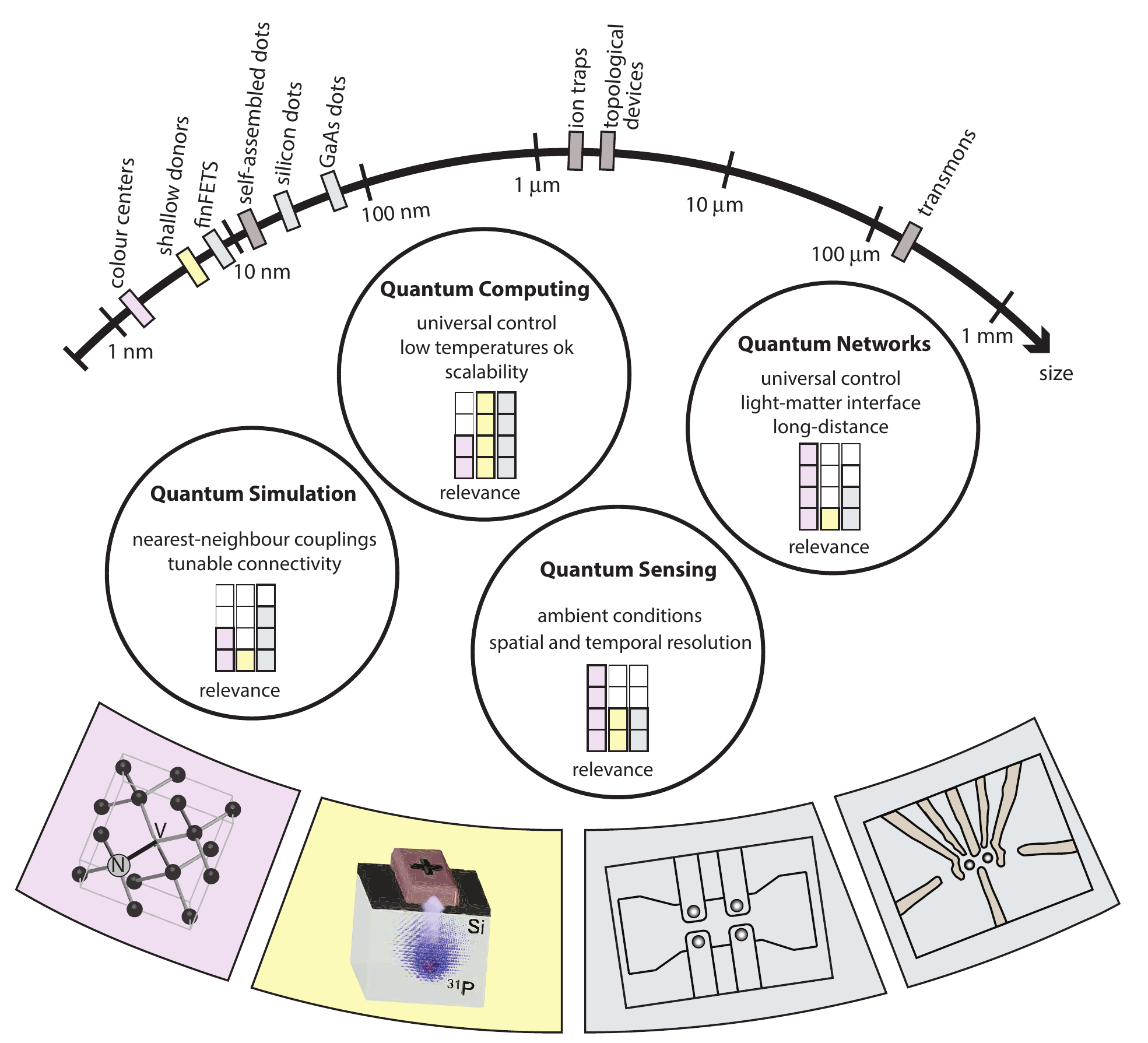}
	\caption{
		\textbf{Semiconductor qubits categorized by size and application}. 
		Top graphic: various solid-state qubit implementations arranged by typical footprint. Middle graphic: Four areas of quantum applications, with different requirements and their suitability to a particular physical implementation (denoted by colour according to bottom graphic). For the purposes of quantifying a system for quantum simulation (as opposed to potential for fault-tolerant quantum computation), we have considered the ability to deterministically place and control the interactions of a set of quantum objects (spin, charge, etc.) to simulate specific physical models, ideally with a simpler architecture and lower effort than that required for a universal quantum computer. Bottom graphic: representative illustrations of the four major qubit categories covered. From left to right, in order of qubit size: color centers, shallow dopants, and gate-patterned quantum dots (silicon-on-insulator accumulation-mode and GaAs depletion mode devices shown here).
	}
	\label{fig1}
\end{figure*}

\section*{Charge in gate-controlled structures}
In a semiconductor structure, the location and motion of an individual electronic charge can be measured with relative ease, and it can as well be used to encode a qubit. Such a charge-based qubit, however, is affected by electrical noise and, for this reason, it is unlikely to achieve the high level of coherence required for quantum computation. On the other hand, the electronic charge degree of freedom is useful for sensing and for reading out of other types of qubit encodings, such as spin qubits. 

In this section we discuss a family of semiconductor-based quantum dots where the confinement of electrons (or holes) is obtained through electrostatic gating in combination with physical or band-structure confinement. The first quantum-dot devices of this family were made from arsenic-based III-V heterostructures (III=Ga,In,Al and V=As) typically grown by molecular-beam epitaxy\cite{reed1988}. A particularly successful approach relied on the use of modulation-doped GaAs/AlGaAs heterostructures hosting a high-mobility two-dimensional electron gas (2DEG). In this case, vertical confinement (along the growth direction) occurs at the GaAs/AlGaAs heterojunction, due to a conduction-band step of a few hundred meV. Lateral confinement is obtained by means of metallic gate electrodes fabricated on the semiconductor surface, which deplete the 2DEG to form a small isolated puddle of electrons.

Because of its versatility, this approach is widely used to realize single and multiple quantum dots with some stability against environmental charge noise. The key of this success lies in a low defect density of the epitaxially-grown heterostructures as well as in the relatively low effective mass of electrons in GaAs, which favors quantum confinement. Moreover, because the 2DEG lies typically 50-100 nm below the surface, the quantum-dot confinement potential is relatively insensitive to surface charges. The occupation of gated quantum dots can be tuned down to the few-electron regime. The exact electron number can be measured by means of a nearby charge sensing device, i.e. another quantum dot or a quantum point contact \cite{elzerman2003}. Increasing the complexity of the gate layout allows going from one, single quantum dot to two or more quantum dots tunnel coupled by gate-tunable interdot barriers (see example in Fig.~\ref{fig1}).

If GaAs-based heterostructures have played a pivotal role in the development of QD devices, much of the research focus is shifting toward silicon-based nanostructures, owing to their potential for spin-based quantum computing. Isotopically purified silicon, enriched in nuclear-spin-free $^{28}$Si, has been shown to provide long spin life times as discussed in the next section. In addition, silicon is an attractive material for large-scale integration. Due to the larger effective masses in silicon, however, the characteristic size of the QDs needs to be smaller than in GaAs, i.e. less then $\sim20$ nm, imposing nanolithographic challenges.
Currently, foundry-fabricated silicon-on-insulator (SOI) transistor technology (incorporating gates over a silicon nanowire, see Fig.~\ref{fig2}(a)), and devices frabricated from high-mobility Si/SiGe strained heterostructure (see Fig.~\ref{fig2}(b)) show promise for industrial mass production, at least on the single- and two-qubit level.

\subsection*{Charge qubit operation and coherence}
A single electron trapped in a pair of adjacent quantum dots can encode a charge-based qubit. The two basis states correspond to the electron sitting either on the left or on the right quantum dot. We label them as $\lvert L \rangle$ and $\lvert R \rangle$, respectively. Figure \ref{fig4}(a) shows the energy level structure of the double QD as a function of the detuning parameter, $\epsilon$, defined as the energy difference between states $\lvert L \rangle$ and $\lvert R \rangle$ in the limit of vanishing coupling. Due to interdot tunneling, the two states hybridize forming bonding and anti-bonding combinations with minimal energy splitting, $2t$, where $t$ is tunneling amplitude. After initializing the qubit in the $\lvert R \rangle$ state, coherent charge oscillations can be induced by a non-adiabatic gate voltage pulse (red) towards $\epsilon \approx 0$ for a controlled amount of time.

Initially, these oscillations were measured in electron transport, manifesting as periodic modulations of the current through the double QD as a function of the pulse duration, $t_p$ \cite{hayashi2003}. Successively, they were detected by measuring the charge occupation of the double QD via a nearby charge sensitive device, such as a quantum point contact or another quantum dot\cite{petersson2010a}, see Fig.~\ref{fig3}(a). The latter approaches eventually enabled a demonstration of full charge qubit functionality with two-axis control \cite{cao2013}. 
More recently, another type of charge-based qubit was experimentally demonstrated in a silicon double QD confining electrons. Such an alternative qubit exploits the multi-valley nature of the silicon conduction band. In a silicon QD, size confinement results in a partial lifting of the six-fold valley degeneracy leaving two lowest energy valleys separated by an energy gap, the so-called valley splitting, that can vary between a few and hundreds of $\mu$eV. In the case of relatively small valley splitting, the QD state can be set in a superposition of the two valley components. Coherent valley-state oscillations \cite{schoenfield2017} and, eventually, a valley qubit with two-axis control \cite{penthorn2019} were demonstrated in a silicon double QD. 
Qubit readout was realized by projecting the valley states onto different charge configurations of the double QD, which were measured with a nearby charge sensor.
While the presented charge qubits allow for extremely fast quantum operation, they are generally sensitive to charge noise, which makes their coherence time rather short and limits the number of consecutive coherent rotations. The detrimental effect of charge noise can be partly mitigated by operating the qubit in a regime where the energy difference between bonding and antibonding states is first-order insensitive to electric field fluctuations. In the case of the valley qubit, this occurs at large detuning, where the coherence time can reach several ns \cite{penthorn2019}. In the case of conventional charge qubits, which rely on a single excess charge shared between two adjacent quantum dots, the sweet spot is at $\epsilon \approx 0$, where the energy difference between bonding and antibonding states is first-order insensitive to charge noise (i.e. to fluctuations in $\epsilon$). 
Petersson {et al.} \cite{petersson2010a} showed a maximum coherence time of about 7 ns at $\epsilon=0$, comparable to that of valley qubits. 

Further improvements in the sweet-spot coherence time can be achieved through material optimization and device engineering. Mi {et al.} \cite{mi2017} reported a coherence time of 0.4 $\mu$s in a silicon-based double QD fabricated from a strained Si/SiGe heterostructure similar to the one shown in Fig. 2(b).  A similarly long coherence time was obtained by Scarlino {et al.} \cite{scarlino2019} using a double QD defined in a GaAs/AlGaAs heterostructure.

While charge makes it difficult to engineer long-coherence qubits, it provides an easily accessible degree of freedom for readout.
Indeed, most highly coherent spin qubits are read out by a process based on spin-to-charge conversion precisely for this reason (see Box 1). These remarkable advances have been instrumental in establishing a strong quantum mechanical coupling between a semiconductor charge qubit and the photon field of a superconducting microwave resonator, opening the door to circuit quantum electrodynamics with semiconductor-superconductor systems \cite{Mi2018, Samkharadze2018, landig2018}.

\subsection*{Applications}
\subsubsection*{Quantum Sensing}
Charge qubits are perhaps most useful as a tool to map and sense other qubit encodings into; via techniques such as spin-to-charge conversion, Pauli spin blockade, and the quantum capacitance of bonding-antibonding orbitals (see Box 1 and Fig.~\ref{fig3}(a-d)), the location of a charge (or its presence or absence) often maps directly onto a spin, a valley or a hybrid spin-charge degree of freedom. 
Another sensing application of charge qubits results, conversely, from their sensitivity to noise; they can be used to detect noise fluctuations with high bandwidth.

\subsubsection*{Quantum Simulation}
A primary characteristic of gate-controlled quantum dots is their versatility; the arrangement of gate patterns, especially with a two-dimensional lateral degree of freedom leads to many ways to spatially engineer a specific Hamiltonian. However, the maintenance of charge coherence over a large array is difficult, though noisy intermediate scale quantum (NISQ) applications for simulation may be envisioned, as for superconducting qubits. These arrays may be more suited to spin qubits (addressed in the next section), with charge qubits useful as readout ancillae.

\subsubsection*{Quantum Computation}
The sensitivity of charge-based states to electrical gate noise and environmental charge noise may rule them out as qubits in a future quantum computer. However, they will likely find a role as readout tools, or in the initialization stage, due to the ease of charge shuttling and charge manipulation.

\subsubsection*{Quantum Communication}
The benefits of readout and coupling to resonators have facilitated the coupling of charge qubits over millimeter-long distances on a chip, via superconducting resonators. By hybridising the charge-spin degrees of freedom, coherent coupling between spins can be achieved. Extending these interactions to form light-matter networks (Fig.~\ref{fig5}(a)) could facilitate quantum communication, especially via conversion of a resonator photon into the telecom wavelengths, an area of ongoing research.

\section*{Spin in gate-controlled structures}

Semiconductors provide another natural implementation of a two-level system utilising the spin degree of freedom\cite{Loss1998}. While individual electronic (or nuclear) spins of dopants embedded in semiconductors can serve as ultracoherent quantum bits (see next section), in this section we focus on gate-controlled quantum dots in which slow and fast gate voltages are used to confine and manipulate electron or hole spins. Due to the need to overcome the thermal energy and suppress phonon-assisted excitations, qubit operation is performed at millikelvin temperatures, even though it was recently shown that spin qubits can remain functional above 1 K \cite{Yang2020,petit2020}. In few-electron  quantum dots, the electron spin has shown long coherence times \cite{Veldhorst2014}; in materials that are poor in spinful isotopes, such as silicon,  isotopic purification has enabled researchers to achieve dephasing times, $T_2^*$, routinely exceeding tens of  microseconds (and, occasionally, even up to 120 $\mu$s \cite{Veldhorst2014}), i.e. much longer than manipulation times\cite{Yoneda2018}.

\subsection*{Qubit encodings}
Many spin-based qubit encodings exist, and in the last decade spin qubits involving more than a single electron have been shown to offer advantages such as electrical control and decoherence-free subspaces. However, tradeoffs in terms of increased operational complexity and fabrication overhead can be involved. Here we describe a few qubit encodings of particular interest.
\begin{itemize}
\item The single-spin (Loss-DiVincenzo) qubit\cite{Loss1998}: In the presence of a static magnetic field, the Zeeman-split $\ket{\uparrow},\ket{\downarrow}$ spin states of an unpaired electron confined in a QD form a paradigmatic qubit encoding. A time-dependent modulation of a magnetic field perpendicular to the one that creates the static energy splitting provides a means to execute coherent single-qubit operations\cite{Koppens2006,Veldhorst2014}.  Exchange-based two-qubit gates can be executed by controlling the wavefunction overlap associated with two spins\cite{Brunner2011,Veldhorst2015}. In addition to singly-occupied QDs, larger (odd) occupation numbers have also been explored\cite{West2018,Higg2014,Leon2020}, and may offer advantages in terms of electrical screening or as intermediate-range couplers\cite{Malinowski2019}. 
\item The singlet-triplet qubit: Encoded in the singlet ($\ket{S}=(\ket{\uparrow\downarrow}-\ket{\downarrow\uparrow})/\sqrt{2}$) and unpolarized triplet ($\ket{T_0}=(\ket{\uparrow\downarrow}+\ket{\downarrow\uparrow})/\sqrt{2}$) states of two electrons in a double quantum dot, here the qubit splitting can be set and controlled by gate voltages\cite{Petta2005}. Coherent rotations can be achieved via a magnetic field gradient between the dots, with the second axis about the Bloch sphere provided by the finite exchange energy. The qubit can be operated in the symmetric regime\cite{Reed2016,MalinowskiSym2017,bertrand2015}, and---with a magnetic quantum number  $m_\mathrm{s}=0$ for both qubit states---is robust to global magnetic field fluctuations.
\item The exchange-only qubit: Utilizing three electrons in a triple quantum dot\cite{Medford2013,Eng2015}, this qubit provides two axes of rotation via the gate-voltage controlled (or resonantly driven) exchange interaction, mitigating the need for a magnetic field gradient, at the cost of more complex operation and a heightened sensitivity to charge noise. For sufficiently large interdot tunneling, these devices can be operated in a rotating frame as resonant exchange (RX) qubits\cite{MalinowskiSym2017,Medford2013,landig2018}.
\item The charge-spin hybrid qubit\cite{Kim2014}: Demonstrated as an all-electrical, double-dot, three-electron qubit with fast rotations ($\pi$-rotations within 100~ps), the hybrid qubit combines advantages from its charge-like (speed) and spin-like (increased coherence) nature.
\item Multi-dot spin qubits: Generalizing the singlet-triplet and exchange-only qubits to collective spin states of multi-dot systems, the quadrupolar exchange-only (QUEX\cite{Russ2018}) qubit and the exchange-only singlet-only (XOSO\cite{Sala2020}) qubit were recently proposed. Both of them exploit a decoherence-free subspace at an extended charge sweet-spot, and potentially can be operated fast, either resonantly or via electrical voltages.
\end{itemize}

Apart from the electron spin, confined holes\cite{Vukusic2017,Maurand2016,Hendrickx2019}, i.e. missing valence band electrons, provide for the realisation of another kind of spin qubit, one where the contact hyperfine interaction is suppressed owing to the p-wave symmetry of valence-band states. 
Here, intrinsic spin-orbit interaction is sufficiently strong to enable hole spin rotations driven by purely electrical means.

\begin{figure*}[!htbp]
	\includegraphics[width=\textwidth]{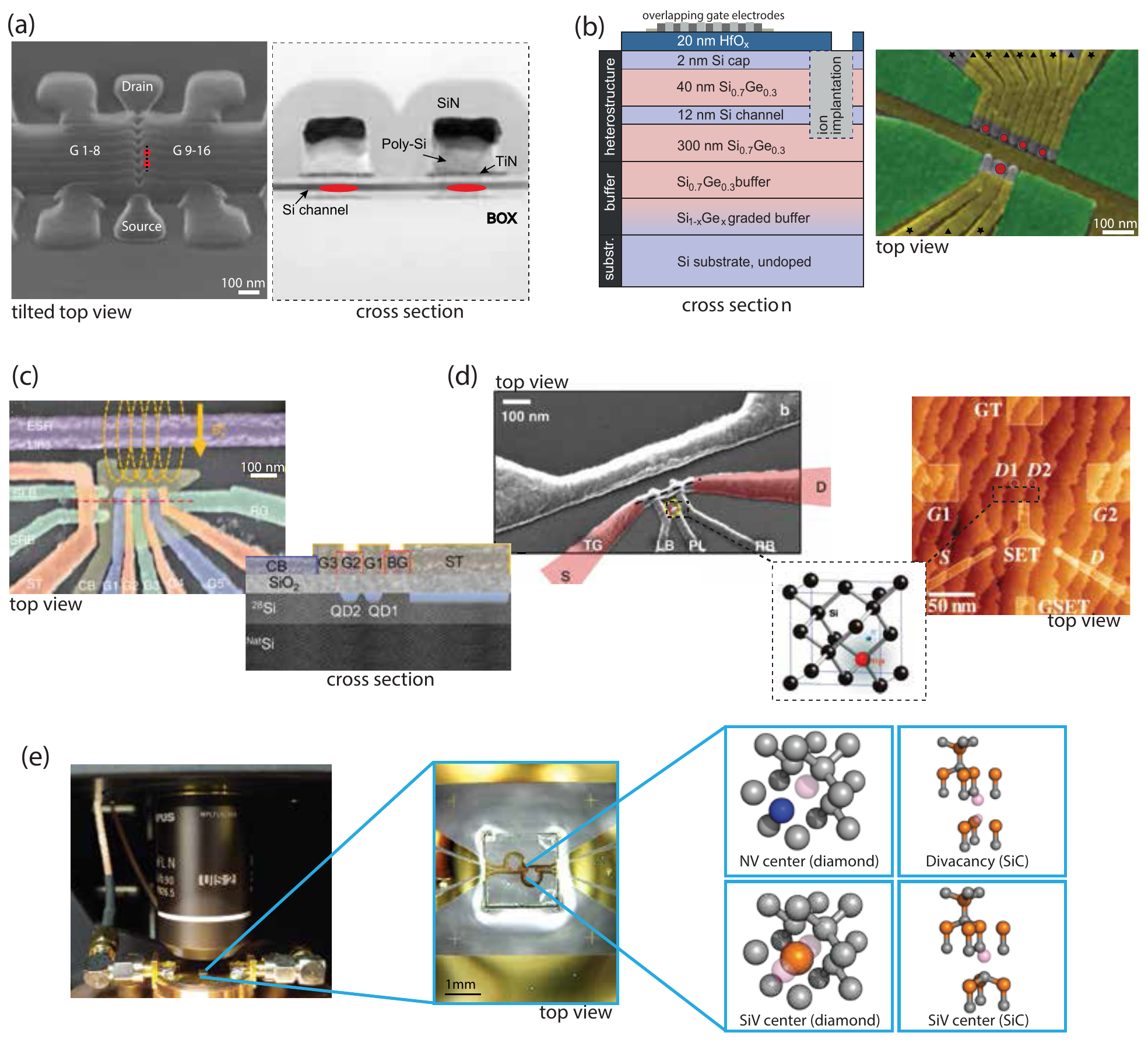}
	\caption{
		\textbf{Materials systems hosting semiconductor qubits}. 
		(a) Quantum dots in nanowires. Confinement is provided in two dimensions by the structure of the nanowire itself (materials may be germanium, InAs, silicon, etc.), while gate electrodes below or above the nanowire create a confining potential along the nanowire. Shown here are finFET-like gates overlapping a silicon nanowire in a silicon-on-insulator device\cite{Crippa2019}, in cross section and top view. Images courtesy of Louis Hutin (CEA-Leti).  (b) Heterostructured materials, typically grown by molecular beam epitaxy or chemical vapour deposition. Here, a silicon-germanium heterostructure is shown in cross section, with a graded buffer for relieving strain  and using bandstructure engineering to create quantum wells\cite{Volk2019a}. Another well-known example is GaAs. Gates are patterned on top (shown in top view) to confine carriers within the plane of the quantum well; two or three layers of overlapping gates (star and triangle symbols indicate different layers, in addition to the green screening layer) can greatly enhance confinement and control\cite{Zajac2016}. (c) A planar silicon structure can be used, with overlapping aluminium gates confining carriers to form qubit dots or SETs\cite{Veldhorst2015}. (d) Shallow dopants can be implanted into a suitable crystal (here phosphorus, a dopant with an extra electron, into silicon, as shown in the inset) with a gate above the implantation location for tuning of the dopant's chemical potential. Gates (here, for an SET charge sensor) can be patterned on top of the crystal as for gate-patterned quantum dots, to read out the dopant charge or spin state\cite{pla2012}, or couple it with a quantum dot. Another approach (right panel) involves engineering the silicon surface atom-by-atom, placing dopant atoms for qubits, SET and gates. (e) Optically active point defects can be addressed by a combination of confocal microscopy and microwave manipulation, left. Substitutional defects and vacancies are common motifs in these defects, such as the nitrogen-vacancy and silicon-vacancy centers in diamond and the divacancy and silicon-vacancy complexes in silicon carbide (right).
	}
	\label{fig2}
\end{figure*}

\subsection*{Initialisation, Readout and Manipulation}
Usually, qubit initialization is not performed by thermalization to the spin ground state since spin relaxation times can be in the range of seconds. Therefore, faster methods are used, such as hot-spot thermalization\cite{Yang2020} or initialising by aligning a particular level of the two level system to a reservoir (spin-selective tunneling)\cite{Veldhorst2015}. 
For the singlet-triplet qubit, the singlet state is most easily initialized in the (2,0) charge state, whereas adiabatic separation into the (1,1) state would initialize the ground state in the nuclear basis ($|\uparrow \downarrow \rangle$ or $|\downarrow \uparrow \rangle$)\cite{Petta2005}. 
Currently, both hole and electron spins can be initialised with fidelities above 99\%.

\begin{figure*}
	\includegraphics[width=\textwidth]{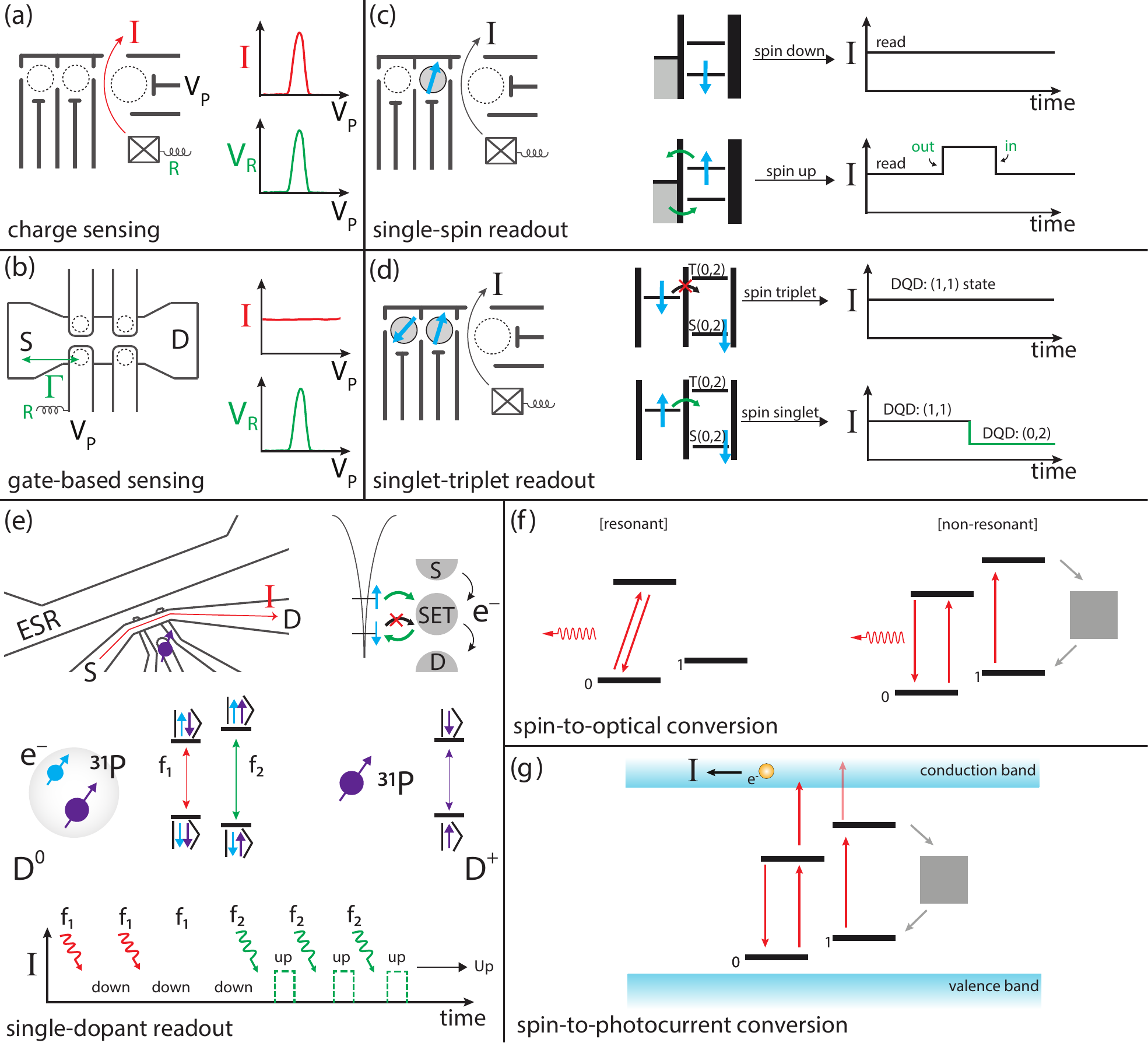}
	\caption{
		\textbf{Readout techniques for semiconductor qubits}. 
		See Box 1 for further details of each readout method. Proximal (a) and gate-integrated (b) charge sensors for readout of charge and spin qubits in gate-controlled semiconductor quantum dots.
		$I$ stands for a DC current flowing through the sensor, which is sensitive to charge rearrangments in its environment. 
		Alternatively, $R$ stands for reflectometry signal (typically in the radio-frequency range), used to read out the sensor with high bandwidth (about 10 MHz). 
		Readout mechanisms for two common qubit encodings, the single-spin qubit ((c), ``Elzerman readout'') and the singlet-triplet spin qubit ((d), Pauli spin blockade). (e) Readout of shallow dopants using a single-electron transistor (SET) charge sensor. (f) and (g) Spin-to-optical and spin-to-photocurrent readout techniques demonstrated for color centers.
	}
	\label{fig3}
\end{figure*}

\begin{figure*}
	\includegraphics[width=\textwidth]{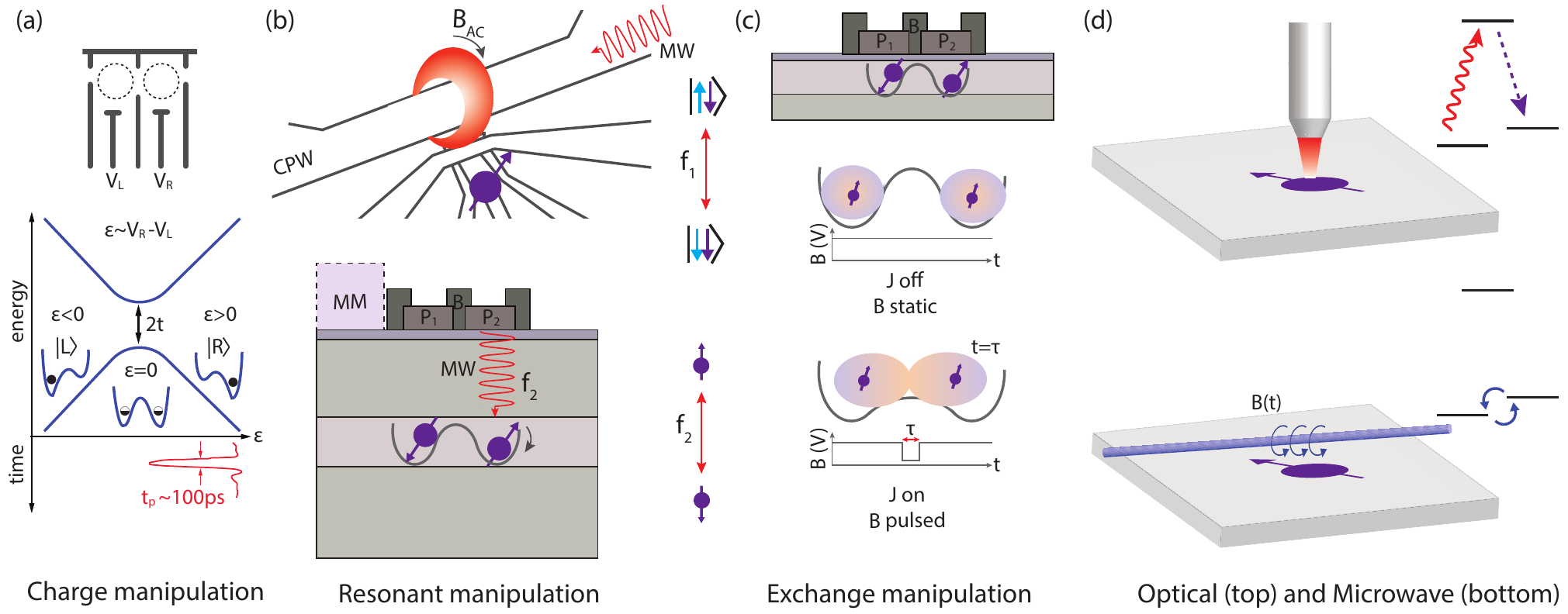}
	\caption{
		\textbf{Manipulation methods for semiconducting qubits}. 
		Depending on the material system and qubit encoding, a host of manipulation methods are available. (a) For charge, voltage pulses applied to gate electrodes can reconfigure potentials in short timescales. Voltages ($V_\mathrm{L}$, $V_\mathrm{R}$) applied to the gate electrodes can move an electron to the left or right, or delocalise it over the two quantum dots creating bonding and antibonding orbitals, by tilting the potential over picosecond timescales (bottom panel). (b) Resonant manipulation can be performed via precisely timed AC fields, which can be magnetic, directly driving the spin via a microwave stripline (top) or electric, shaking the position of the spin in a natural\cite{Nadj-Perge2010} or synthetic\cite{Yoneda2018} (micromagnet-generated, see square labelled ``MM'', bottom panel) spin-orbit fields. (c) The Heisenberg exchange interaction can be used to rotate spins by a specified amount, by finely tuning the degree of wavefunction overlap\cite{Petta2005}. This can be done on fast timescales ($\sim$ ps to ns), using gate voltage pulses; for example, by lowering the barrier gate (B, top panel) in the case of symmetric operation\cite{MalinowskiSym2017}. For three spins instead of two, exchange-based gates can provide for universal one- and two-qubit operations\cite{Medford2013,DiVincenzo2000}. (d) Optical pumping provides a convenient route to initializing optically-active defects in a specific state, while resonant microwave fields allow for coherently controlling different spin projections.
	}
	\label{fig4}
\end{figure*}

Despite the multiple ways of encoding spin qubits described above, the primary readout method is based on spin-to-charge conversion\cite{elzerman2004}. 
While the magnetic moment of a single spin is exceedingly small (of order of the Bohr magneton, $\sim$57.8~$\mu$eV/T) and its direct detection correspondingly difficult, the detection of small displacements of charge has been perfected over the years. 
Techniques based on a local charge detector\cite{Petta2005}, radio-frequency reflectometry\cite{Petersson2010,Barthel2009} or dispersive readout via a resonator\cite{Zheng2019,Gonzalez-Zalba2015} have been developed (see Fig.~\ref{fig3}(a,b)). To implement spin-to-charge conversion, a charge displacement between dots, or between a quantum dot and a reservoir, is engineered to be dependent on the spin-qubit state (see Box 1).
A separate charge detector can be employed for this measurement; this can be a constriction in a nearby 2DEG, called a quantum point contact\cite{Reilly2007}, a quantum dot\cite{MalinowskiSym2017}, a single-electron transistor (SET)\cite{Veldhorst2014}, or a direct measurement of charge oscillation by detecting the resulting gate-capacitance changes\cite{Petersson2010,Gonzalez-Zalba2015,Crippa2019}. Charge detectors in conjunction with spin-to-charge conversion and radio-frequency reflectometry have yielded single-shot spin-state measurements in microseconds\cite{Zheng2019}, while readout fidelities of 99.8\% have been reported\cite{pla2013,Yoneda2018}.

Each encoded qubit requires methods for individual rotations about two axes of the Bloch sphere, as well as entanglement between two neighbouring qubits, in order to form a set of universal quantum gates. Typically, for gate-controlled spin qubits, these logic gates are achieved by two methods. 

Directly engineered magnetic fields, turned on and off for short timescales using an coplanar stripline fabricated adjacent to the spin qubit, can be used to performing single-spin rotations via electron spin resonance (ESR)\cite{Koppens2006,Veldhorst2014}. This method involves using resonant microwave pulses that match the qubit Larmor frequency and therefore drive rotations between different energy levels.

Fast gate-voltage pulses can be used to move the electron or hole wavefunction. Baseband pulses are used to switch on and off the Heisenberg exchange interaction (Fig. 4c) to control singlet-triplet and exchange-only qubits. Microwave modulated pulses are used to oscillate the electron wavefunction to obtain an effective time-varying magnetic field via a natural\cite{nowack2007coherent,Nadj-Perge2010,Corna2018,Crippa2018} or synthetic\cite{pioro2008electrically,Kawakami2014,Zajac2018} spin-orbit field (Fig. 4b, bottom). The result is a form of spin resonance that involves an electric dipole transition, thus called Electric Dipole Spin Resonance (EDSR).

\subsection*{Material systems}
There are several material systems suitable for gate-controlled spin qubits, with specific advantages and disadvantages for particular applications. These include engineered heterostructures (where charge carriers are strongly confined along the growth directions), nanowires (which naturally provide confinement in two directions) as well as planar semiconductor platforms (see Fig.~\ref{fig2}). 

Starting in 2005, the first experiments demonstrating spin qubits were reported in GaAs/AlGaAs heterostructures\cite{Petta2005}, where the 2DEG formed in the GaAs layer is depleted by negative gate electrodes to trap individual electrons for qubit operations. The GaAs platform benefits from a relative simplicity of fabrication, as well as some favourable electronic properties, such as a single conduction band valley and a small effective mass leading to less stringent lithographic constraints. On the other hand, the totality of the atoms in the lattice carry a nonzero nuclear spin, making hyperfine interaction a significant source of decoherence and leading to intrinsic inhomogeneous dephasing times of $T_2^*\approx$ 10~ns. However, more than a decade of technological improvements in GaAs, namely the development of dynamic nuclear polarization\cite{Foletti2009}, dynamical decoupling sequences\cite{Malinowski2017} as well as nuclear field distribution narrowing have enabled millisecond-long coherence times and single qubit control with a fidelity of 99.5\%\cite{Cerfontaine2019}. 
Additionally, piezoelectricity and spin-orbit coupling are other concerns in this material system, while the direct band gap could be useful for spin-photon conversion. Currently, research continues on GaAs heterostructure-based devices for proof-of-concept multi-spin quantum devices\cite{Mortemousque2018,Malinowski2019}, spin transfer demonstrations\cite{Kandel2019} and quantum simulators\cite{Dehollain2020}.

Since the demonstration of the first silicon spin qubits in 2012\cite{pla2012,Maune2012}, research focus has moved to this low-nuclear-spin material system. Here, electrons (or holes) are confined in silicon MOS devices with either planar\cite{Veldhorst2014} or nanowire\cite{Maurand2016} structures, or in devices based on silicon-germanium (Si/SiGe) heterostructures\cite{Kim2014,Kawakami2014,Zajac2016,Zajac2018}. Dopant atoms in the silicon host can also be used as qubits\cite{pla2012,pla2013}, as discussed in Section 4. 

Natural silicon contains only 4.7\% of $^{29}$Si, the only stable isotope bearing nuclear spin; this can be reduced to ppm concentrations by isotopic purification\cite{Itoh2014}. Silicon also has a weaker spin-orbit interaction than GaAs, InAs, and InSb, and is a material compatible with powerful foundry fabrication in the microelectronics industry.

Currently, silicon spin qubits are some of the most coherent spin qubits, with gate-controlled implementations showing a (dynamically decoupled) coherence time up to 28~ms\cite{Veldhorst2014}. However, some challenges remain\cite{Zwanenburg2013}. In silicon, devices need to be smaller compared to GaAs devices, due to the larger effective mass of electrons, and fabrication results are not yet as reproducible as in GaAs. As a result of valley degeneracy, low-lying leakage states are possible that may be thermally populated even at low temperature\cite{Zwanenburg2013}. In future years, research on silicon qubits must focus on the scalability considerations due to the fact that valley splitting is affected by unavoidable fabrication-related defects, inhomogeneities and step edges in the nanowire, interface or heterostructure.

Germanium, another group-IV semiconductor and the material of the first transistor, has recently been used to make SiGe/Ge/SiGe quantum-well heterostructures confining high-mobility hole gases and quantum dots that can be used to encode spin qubits\cite{Vukusic2017,Watzinger2018,Hendrickx2019}. Due to the inherent presence of a sizeable spin-orbit coupling in the valence band, hole qubits can be manipulated by means of EDSR without the need for local micromagnets. The characteristic manipulation times demonstrated so far ($\sim$10 ns) are much shorter than those achieved for electrons, which partly compensates for the shorter hole coherence times.

\begin{table*}[!htbp]
	\centering
	\begin{tabular}{|p{\textwidth}|}
		\hline
		\textbf{Box 1. Spin-to-Charge Conversion: A workhorse for semiconductor qubits}\\
		\hline
		
		For spin qubits in semiconductors, measurements of the spin state are typically made through projection into another degree of freedom such as charge, which is more easily accessible by transport and charge-sensing measurements. Spin-to-charge conversion has become the dominant way to read out quantum dot spin states. Here, we illustrate this principle for a few different qubit implementations.
		\begin{itemize}
			\item Figure 3(c) shows a sketch of spin-to-charge conversion based on energy selection. A single spin in a quantum dot is capacitively coupled to a sensor and tunnel coupled to a reservoir. After spin manipulation, the dot energy level is tuned such that the Fermi reservoir lies between the two Zeeman-split spin states ($\uparrow$ and $\downarrow$). If the dot is in state $\downarrow$, the Coulomb blockade prevents the electron from entering the reservoir, whose energy levels are completely filled at that energy, and there is no sensor change due to charge rearrangement.
			For a state $\uparrow$, the electron can tunnel out of the quantum dot and into empty states of the reservoir located above the Fermi energy, leading to a measurable charge change until a new electron tunnelling in reinitializes the qubit to its ground state. 
			\item For a double-dot singlet-triplet qubit, capacitively coupled to a sensor, the conversion process makes use of spin-to-charge conversion (Fig.~\ref{fig3}(d)). Here, the two levels are the singlet ($\ket{S}$, typically in the (0,2) charge state) and the (1,1) triplet state ($\ket{T_0}$). After spin manipulation in the (1,1) charge state (with one electron in each quantum dot), the dot levels are tilted to favour the (0,2) charge state. 
			If the two spins are singlet correlated (antisymmetric spin states), the left electron can tunnel to the same orbital occupied by the right electron, giving rise to a charge change detected by the sensor. If the two spins are parallel (symmetric spin states), Pauli exclusion prevents such a tunneling, unless higher lying orbitals in the right dot can be accessed.   
			This conditional tunneling process arising from spin blockade is known as spin-to-charge conversion.
			\item Shallow dopants also utilise the process outlined for single-spin state conversion to charge. However, in all qubit demonstrations to date, a single-electron transistor (SET) acts both as the charge sensor and as the charge reservoir for spin-to-charge conversion. The dopant energy levels which form the two-level system for the ionised nuclear spin and the electron spin qubit are shown in Fig. 3(e), and transitions are addressable using distinct frequencies ($f_1$ and $f_2$ in the figure). After manipulation, as described for the first case, the dopant energy levels are aligned with the SET, and conditional tunneling is used to both measure and reinitialise the qubit.
			\item In color centers, spin-to-charge conversion can be used either as a direct readout scheme (via photocurrent), or as a contrast mechanism. Both schemes rely on state-dependent photoionization; in the case of the NV center in diamond, shelving into a metastable singlet state protects the $m_s=\pm1$ states from ionization. The resulting photocurrent can be measured directly with local electrodes, or the ionization into the spectrally-distinct NV$^0$ charge state may be used to increase contrast in traditional photoluminescence-based readout schemes.
		\end{itemize}
		
		\\
		
		\hline
	\end{tabular} 
\end{table*}

\subsection*{Applications}
\subsubsection*{Quantum Sensing}
Quantum sensing is well suited to spin qubits since they are sensitive magnetometers, as well as excellent detectors for charge noise at appropriate operating points. A gate-tuned exchange interaction, for example, is finely dependent on electrical fields and can be used to detect both low and high-frequency electrical signals. 
In the magnetic domain, the nuclear spin bath and its diffusion constant has been measured using the dephasing of a spin qubit in GaAs\cite{Malinowsky2017Spec}. However, impurity atoms such as shallow donors or NV centers are currently more suited to technologies such as scanning-probe magnetometers. In future, flip-chip technology\cite{Tahan2020}, already available for gate-controlled superconducting qubits, could revolutionise mobile sensing applications for gate-controlled spin qubits. In addition, through fast spin-to-charge conversion techniques, spin qubits can be local probes for other physical systems that can also be implemented in semiconductors such as nanowires and high-mobility 2DEGs. When these systems interact in specific ways with spin, they can be probed in a range of bandwidths; examples include the Kondo state, quantum Hall edge states and topological entities such as Majorana fermions\cite{Hoffman2016}. However, the requirement for millikelvin temperatures and wiring requirements, combined with the lack of optical addressability, make it difficult for these spin systems to be useful for biological or environmental sensing.

\subsubsection*{Quantum Simulation}
As for charge qubits, gate-controlled quantum dots hosting spin qubits lend themselves to two-dimensional arrays\cite{Mortemousque2018}, which can spatially engineer specific Hamiltonians. The gate-controlled exchange interaction for spin can then be used to form as-desired connections between these array components on GHz timescales (see Fig.~\ref{fig5}(c)). Arrays of spin qubits have lent themselves to simulation experiments, such as that of a Mott insulator based on the Hubbard model\cite{Hensgens2017} and for probing itinerant magnetism in the Nagaoka regime\cite{Dehollain2020}.

\subsubsection*{Quantum Computation}
Along with superconducting qubits, gate-controlled spin qubits are one of the technologies that present an advanced level of development with regards to coherence, scalability and the ability to make small-scale quantum processors, making quantum computation one of the primary long-term applications of these systems\cite{veldhorst2017silicon,vandersypen2017interfacing}. With greater research focus on the silicon materials system, especially with semiconductor companies and research foundries entering the fray, foundry-fabricated silicon qubits are becoming a reality and demonstrating the scalability of the gate-controlled approach. 
At the same time, SiGe and planar-silicon devices have shown long coherence times, especially impressive when compared to short gating times, yielding gate fidelities above some error correction thresholds. 
Finally, with the demonstration of reliable two-qubit gates\cite{Veldhorst2015,Zajac2018,Watson2018,Hendrickx2019}, all of the DiVincenzo criteria have been satisfied and simple proof-of-principle quantum algorithms\cite{Watson2018} have been implemented.

\subsubsection*{Quantum Communication}
Quantum communication involves non-classical communication and distant entanglement, whether direct or through an intermediary classical or quantum system. For spin qubits, direct entanglement is a difficult goal to achieve, since the long coherence times of spins are due to their isolation from the environment, and spin-spin interactions are intrinsically short-range. However, intermediary systems have been used to extend the range of this interaction over longer distances; one way is to use an exchange coupling mediated via a large multielectron dot\cite{Malinowski2019}. An active area of research for long-distance coupling is a spin-cQED architecture, recently reviewed\cite{Burkard2020}, where a spin-spin or spin-photon coupling is achieved via or to a superconducting circuit\cite{Borjans2020,Mi2018,Samkharadze2018,landig2018}, with potential for conversion to a photon in the telecommunications wavelength. Physical shuttling of electrons\cite{Bertrand2016} using a surface acoustic wave or finely-tuned and timed voltage pulses\cite{Mills2019} has also been demonstrated. Also, as mentioned before, the direct gap in GaAs and other materials could be useful for a direct exciton-photon coupling.

\section*{Dopants in silicon}

Shallow group-V donors in group-IV materials constitute a solid-state analog of the hydrogen atom. For example, phosphorus in silicon (Si:P) possesses a weakly bound electron in a 1s-like orbital wavefunction, and a spin-1/2 nucleus coupled to the electron via Fermi contact hyperfine interaction. The 1s orbital envelope wavefunction has a Bohr radius of about 2~nm: an estimate for this value could be obtained from the hydrogen Bohr radius formula $a_0 = 4\pi\varepsilon_0 \hbar^2/m_e e^2$, replacing the permittivity of vacuum $\varepsilon_0$ with the dielectric constant of silicon, and the electron mass $m_e$ with the effective mass $m^*$ of an electron near the bottom of the silicon conduction band. 

Looking beyond the hydrogenic approximation, the physics of spin and orbital states of shallow donors reveals important details on band structure properties such as valley degeneracy of the conduction band minima, spin-orbit coupling, or valley-orbit coupling. Therefore, at the dawn of modern semiconductor electronics, detailed studies of the quantum properties of dopant atoms in silicon \cite{kohn1955} constituted an important benchmark for band structure theories, which were being developed in the 1950s. It was well known, even back then, that the spins of electrons bound to dopants in silicon possess long-lived quantum states \cite{feher1959}.  

\subsection*{Spin coherence in donor ensembles}

Once quantum computing became a topic of active research, it was therefore natural to imagine using donor spins for quantum information processing. In 1998, Kane presented a visionary proposal to encode quantum information in the nuclear spin of individual $^{31}$P donor atoms in silicon \cite{kane1998}. He warned about the extreme technological challenge in fabricating devices at the single-atom level, but argued that the progress in miniaturization imposed on the semiconductor industry by the pursuit of Moore's law would eventually lead to the capacity to fabricate silicon devices at the scale necessary for quantum computing. Twenty years later, this vision has indeed been materialized.

The renewed interest in the spin coherence of donors in silicon motivated a series of electron spin resonance (ESR) studies on bulk spin ensembles, particularly as a function of isotopic enrichment of $^{28}$Si. Samples with residual $^{29}$Si (with nuclear spin $I=1/2$) concentration of 800 ppm showed electron coherence times $T_{\rm 2e} \approx 60$~ms \cite{tyryshkin2006}, extensible to 10 seconds by suppressing dipole-dipole interactions \cite{tyryshkin2012} and further reducing the $^{29}$Si concentration to 50 ppm. The accuracy of these experiments made them an ideal test bed for advanced theoretical methods to describe spin dephasing caused by a fluctuating nuclear spin bath \cite{witzel2007}. Such highly coherent spins systems enabled the demonstration of a nuclear quantum memory protocol, where the electron spin state is stored in the $^{31}$P nucleus and later retrieved \cite{morton2008}. Operating at high magnetic field (3.4 T) and low temperature (2.9 K) allowed the establishment of genuine quantum entanglement between the electron and the nuclear spin of the P donors \cite{simmons2011}. Using the spin-dependent photoionization of the donors to detect the nuclear spin polarization, donor ensembles in highly enriched $^{28}$Si exhibited extraordinary nuclear coherence times, $T_{\rm 2n}^0 = 3$~minutes in the neutral charge state \cite{steger2012}, and $T_{\rm 2n}^+ = 3$~hours in the ionized state \cite{saeedi2013}, all obtained using XYXY dynamical decoupling.

\subsection*{Single-donor $^{31}$P spin qubits}

A scalable quantum processor requires individually addressable and measurable donors. The fabrication of single-donor devices took two alternative pathways: one based on hydrogen lithography with scanning tunneling microscopy (STM) \cite{obrien2001}, and the other following the industry-standard ion implantation method \cite{jamieson2005}. The first experimental breakthrough in this field was achieved by combining metal-oxide-semiconductor quantum dots \cite{angus2007} with ion-implanted donors in a tightly integrated structure, in which a large quantum dot acts as a tunnel-coupled charge sensor for the donor-bound electron. Using the energy-dependent tunneling method developed earlier in GaAs quantum dots \cite{elzerman2004} unlocked the capability to read out the spin state of a single implanted P atom in single-shot \cite{morello2010}. The same method was later extended to STM-fabricated donor clusters \cite{buch2013}.

Coherent control of the first single-atom electron \cite{pla2012} and nuclear \cite{pla2013} spin qubits in silicon was achieved by integrating an implanted $^{31}$P atom with the readout circuitry and a broadband, on-chip microwave antenna. The first experiments, conducted using a natural Si substrate, yielded spin coherence times in line with the expectations from ensemble experiments, with electron dephasing time $T_{\rm 2e}^* = 55$~ns, and Hahn echo time $T_{\rm 2e}^{\rm H} = 200$~$\mu$s \cite{pla2012}. Remarkably, no ensemble data exist for the coherence time of nuclear spins in natural Si: the single-spin values for the Hahn echo time are $T_{\rm 2n}^0 = 3.5$~ms and $T_{\rm 2n}^0 = 60$~ms in the neutral and ionized state, respectively \cite{pla2013}.

With the introduction of isotopically-enriched $^{28}$Si substrates, the performance of single-donor qubits improved dramatically, with coherence times reaching $T_{\rm 2e} = 0.56$~s and $T_{\rm 2n}^+ = 35.6$~s using dynamical decoupling \cite{muhonen2014}. These record coherence times translate to 1-qubit Clifford gate fidelities, measured by randomized benchmarking, of 99.94\% for the electron \cite{dehollain2016a} and 99.98\% \cite{muhonen2015} for the nucleus. Electron-nuclear entanglement, mediated by the hyperfine interaction, was demonstrated by violating Bell's inequality with a record Bell signal $S = 2.70$ \cite{dehollain2016b}.

\subsection*{Alternative dopant atoms}

Although the majority of research on shallow donors has focused on $^{31}$P in silicon, alternative dopants offer interesting properties. $^{209}$Bi possesses a large nuclear spin $I=9/2$ and a very strong hyperfine interaction $A=1.4$~GHz, which result in an energy level diagram that comprises ``clock transitions'', i.e. pairs of states whose energy splitting is to first-order insensitive to magnetic field noise \cite{wolfowicz2013}. Ensembles of $^{209}$Bi atoms have been integrated with superconducting resonators to obtain the first demonstration of Purcell effect for spins in the solid state \cite{bienfait2016}. Integrating a single $^{123}$Sb donor (nuclear spin $I=7/2$) in a nanoscale device has led to the discovery of nuclear electric resonance, whereby coherent nuclear spin transitions are induced by the electrical modulation of the nuclear quadrupole coupling \cite{asaad2020}. Acceptor atoms such as boron have been recently studied in ensemble experiments using planar microwave cavities, and have shown remarkably long spin coherence times ($T_2^{\rm H} \approx 1$~ms) in strained samples of enriched $^{28}$Si \cite{kobayashi2018}.

\subsection*{Applications}

\subsubsection*{Quantum sensing}

The high spin coherence of donors in enriched $^{28}$Si translates into a high sensitivity to minuscule magnetic field perturbations. Noise spectroscopy on a single $^{31}$P atom revealed a noise floor equivalent to 18 pT/$\sqrt{\mathrm{Hz}}$ \cite{muhonen2014}. However, the need for low-temperature operation and integration with charge readout devices suggests that the magnetic sensing applications of donors are likely to be limited to the detection of materials and structures fabricated directly on top of the chip.

Recently, the idea of sensing strain at the atomic scale has been suggested. Strain is universally adopted in modern ultra-scaled silicon transistors to maximize their electrical performance. Lattice strain could be detected either through its influence on the quadrupolar splitting of heavy group-V donors such as $^{75}$As \cite{franke2015} and $^{123}$Sb \cite{asaad2020}, or through the shift of the optical transition frequency of Er atoms \cite{zhang2019}.

The recent demonstration of full quantum control of a high-spin $^{123}$Sb nucleus \cite{asaad2020} opens exciting perspectives for exploring highly non-classical spin states for quantum sensing. Enhanced quantum sensing methods pioneered in the cold atoms community \cite{pezze2018} include the use of ``Schr{\"o}dinger cat'' states or spin squeezed states. Their metrological usefulness is a topic of active research, as it depends crucially on the nature of the noise that perturbs the system. It is predicted that the key to achieving a metrological advantage from non-classical spin states is the presence of non-Markovian noise \cite{matsuzaki2011}, which is precisely the kind naturally occurring in solid-state spins immersed in a sparse nuclear spin bath\cite{madzik2019}.

\subsubsection*{Quantum simulation}

The atomic size and amenability to atomically-precise placement makes dopants in silicon an appealing platform to embody solid-state quantum simulations of the Hubbard model, where single-site measurement could be obtained by scanning tunneling spectroscopy \cite{salfi2016a}. In the context of digital quantum simulations, it has been shown that the kicked-top model of quantum chaos can provide precious insights into the proliferation of errors due to discrete Trotter steps \cite{sieberer2019}. A recent proposal suggests that a chaotic top could be experimentally demonstrated in the high-spin nucleus of a $^{123}$Sb donor \cite{mourik2018}.

\subsubsection*{Quantum computation}

Dopant spins in silicon are among the most coherent quantum systems in the solid state, making them appealing candidates for quantum information processing, within the most important material for the classical computer industry. Building a scalable quantum processor requires moving beyond single-spin coherence and 1-qubit fidelities, introducing a coupling between the electrons. The natural coupling mechanism is the exchange interaction \cite{kane1998}, arising from overlap of the donor electron wave functions. Strong exchange has been observed in several two-donor devices \cite{dehollain2014,gonzalez2014}, and fast time-resolved exchange oscillations have been achieved using donor-defined quantum dots, with $t_{\sqrt{S}} = 0.8$~ns for a $\sqrt{\mathrm{SWAP}}$ operation \cite{he2019}. The very short range of this interaction, however, poses significant challenges to the layout of a large-scale processor. Therefore, there have been suggestions for spacing out the donors using large interposer quantum dots \cite{srinivasa2015}, dopant spin chains \cite{mohiyaddin2016} or ferromagnetic couplers \cite{trifunovic2013}.

Alternatively, scaling up donor-based quantum computers might be achieved by adopting different qubit encodings, which possess an electric dipole, and use electric dipole-dipole interaction, or coupling the dipole to a microwave resonator, to achieve long-distance coupling. A natural system in which an electric dipole can be dynamically induced is the boron acceptor. Placing B in strained Si creates a noise-resilient system where dipole-dipole coupling can mediate a $\sqrt{\mathrm{SWAP}}$ gate between acceptor spins in 4~ns at 20~nm distance \cite{salfi2016b}. 

A large artificial electric dipole can be induced on a $^{31}$P donor by placing it at a vertical distance $\approx 15$~nm from a Si/SiO$_2$ interface, and applying a vertical electric field to displace the donor-bound electron towards the interface. This introduces a strong electrical modulation of the electron-nuclear hyperfine coupling, which can be used to electrically drive coherent transitions between the $|\Uparrow\downarrow\rangle \leftrightarrow |\Downarrow\uparrow\rangle$ ``flip-flop'' states. A pair of flip-flop qubits could perform $\sqrt{\mathrm{SWAP}}$ operations in 40~ns at a distance of 200~nm \cite{tosi2017}. A hybrid donor-dot system can also be used in a two-electron configuration to form a singlet-triplet qubit. The magnetic field gradient $\Delta B_z$ is set to a fixed and large value by the presence of the hyperfine coupling in the donor-bound electron, allowing fast exchange oscillations with $t_{\sqrt{S}} = 4$ns \cite{harvey2017}.

In addition to the coupling methods discussed above, some proposals provide detailed prescriptions for the layout and operation of large-scale donor-based quantum computers, where multi-qubit gates mediated by magnetic dipole coupling \cite{hill2015}, potentially provided by a moving probe \cite{ogorman2016}, or shuttling electrons across quantum dots coupled to sub-surface donors \cite{pica2016}. For donor encodings that possess an electric dipole, there is also the option of achieving long-distance coupling mediated by microwave photons in superconducting resonators \cite{tosi2017,salfi2016b}.

\subsubsection*{Quantum communications}

The indirect bandgap of silicon is normally an obstacle for the establishment of optical interconnects between shallow donors, since the decay of an optical excitation typically results in an Auger process, where the energy of the absorbed photon is imparted to the donor-bound electron instead of being re-emitted. More promising systems for spin-optical interface are group-VI donors such as $^{77}$Se, which have much deeper binding potentials than shallow group-V atoms, and possess optical resonances in the mid-infrared, associated with long-lived spin states and magnetic clock transitions\cite{morse2017}. Alternatively, rare-earth atoms such as erbium exhibit spin-dependent optical transitions at telecommunication wavelength, and have been integrated with nanoscale transistors for electrical readout of the optical excitation \cite{Yin2013}.

\begin{table}[]
\resizebox{\textwidth}{!}{%
\begin{tabular}{|c|c|c|c|c|c|c|}
\hline
Qubit type & \multicolumn{2}{c|}{Characteristic timescales (s)} & \multicolumn{2}{c|}{Quantum Computation} & \multicolumn{2}{c|}{Quantum Sensing}  \\ \hline
& $T_\mathrm{1}$           & $T_\mathrm{2}$            & Single-qubit gate time     & Single-qubit fidelity &Quantity       & Sensitivity             \\ \hline
Gated charge   &    30~ns\cite{Kim2015a}    &  7~ns\cite{petersson2010a}             &      $\sim$0.1~ns\cite{cao2013}               &   86\%\cite{Kim2015a}        & Charge   & $\mathrm{\sim 10^{-4}~ e/\sqrt(Hz)}$ at 1~Hz \cite{petersson2010a}       \\ \hline
Gated spin     &    57~s\cite{Camenzind2018}            & 28~ms\cite{Veldhorst2014}        &     0.25~ns\cite{Yoneda2018}                 &  99.96\%$^\ast$\cite{yang2019RB}     &     Magnetic field gradients               &    50~pT$/\sqrt{\mathrm{Hz}}$\cite{Taylor2008,Veldhorst2014}                                \\ \hline
Shallow dopants (electron) &  $> 1$~h\cite{feher1959} (ens), 10~s\cite{Tenberg2019}    &  10~s\cite{tyryshkin2012} (ens), 0.56~s\cite{muhonen2014} &  $\sim 100$~ns\cite{pla2012}   &  99.94\%$^\ast$\cite{dehollain2016b}  &    Magnetic field (AC)                  &    18~pT$/\sqrt{\mathrm{Hz}}$\cite{muhonen2014}                         \\ \hline
Shallow dopants (nucleus) &     $>$~days\cite{pla2013}     &  3~h\cite{saeedi2013} (ens), 35.6~s\cite{muhonen2014}    &   $\sim 20~\mu$s\cite{pla2013}    &   99.98\%$^\ast$\cite{muhonen2015}  &    Magnetic field  (AC)                &     2~nT$/\sqrt{\mathrm{Hz}}$\cite{muhonen2014}                        \\ \hline
\multirow{4}{*}{Color centers}   &  \multirow{4}{*}{>1~h\cite{Abobeih2018a}}      &    \multirow{4}{*}{1s\cite{Abobeih2018a}}          &     \multirow{4}{*}{<20~ns\cite{Robledo2011}}                  &  \multirow{4}{*}{99.995\%$^\ast$\cite{Rong2015}}  &   Magnetic field (DC)             &  50~pT/$\sqrt{\mathrm{Hz}}$\cite{Schloss2018} (ens), 500~nT/$\sqrt{\mathrm{Hz}}$\cite{Liu2019}                                         \\ \cline{6-7}
   &  &   &    &    &   Magnetic field (AC)             &  32~pT/$\sqrt{\mathrm{Hz}}$\cite{Glenn2018a} (ens), 4.3~nT/$\sqrt{\mathrm{Hz}}$\cite{Balasubramanian2009}                                         \\ \cline{6-7}
   &                &             &                       &    &   Temperature                &  100~mK/$\sqrt{\mathrm{Hz}}$ \cite{Anisimov2016,Nguyen2018a}
			\\ \cline{6-7}
   &                &             &                       &    &   Electric Field               &  $10^{-5}~\mathrm{V\cdot \mu m^{-1}/\sqrt(Hz)}$ \cite{Michl2019}
			\\ \hline
\end{tabular}%
}
\caption{
\textbf{Current state of the art for semiconductor qubits}. 
We provide the best known values (at the time of writing) for key single-qubit timescales: the relaxation time, $T_\mathrm{1}$, and the maximum coherence time, $T_\mathrm{2}$, using dynamical decoupling. For quantum computation, we quote the typical single-qubit gate time and the highest single-qubit gate fidelity; an asterisk denotes gate fidelities assessed using randomised benchmarking, and (ens) indicates values obtained in bulk spin ensembles. State-of-the-art experiments for quantum sensing are also highlighted, in terms of the quantity sensed and the sensitivity achieved. 
For gated charge, the reported time scales refer to experiments on semiconductor double quantum dots operated as charge qubits. Yet, much longer sweet-spot coherence times (around 0.4~$\mathrm{\mu}$s) were recently found in double quantum dots coupled to a superconducting resonator\cite{mi2017,scarlino2019}. We also note that charge sensitivity is frequency dependent due to the $1/f^\gamma$ character of electrical noise ($0.7 \lesssim \gamma \lesssim 1.4$ \cite{Connors2019, Mi2018b, Dial2013}), and it can be measured in different device setups. For example, measurements on singlet-triplet qubits yielded a spectral density of $2 \times 10^{-8} e/\sqrt{Hz}$ at 1 MHz \cite{Dial2013}. When scaled down to 1 Hz this value is consistent with the one given in the table.}
\end{table}

\section*{Optically-Addressable Quantum Defects}
Optically-addressable quantum defects are point defects in a lattice where a spin degree of freedom is coupled to one or more optical transitions. This spin-photon interface allows for the combination of two powerful toolkits: spin resonance techniques for manipulating the spin and its interactions with its environment, and single molecule microscopy techniques for addressing individual quantum defects. There has been intense interest over the past two decades in developing quantum defects for a variety of technologies, particularly focused on color centers in diamond such as the nitrogen vacancy (NV) center. Such defects have been deployed in a number of quantum applications, including spin-spin entanglement \cite{Dutt2007,Dolde2013,Neumann2010,Bradley2019}, nanoscale and quantum sensing \cite{Maze2008,Taylor2008,Lovchinsky2016,Aslam2017,Boss2017,Schmitt2017a}, and remote entanglement and quantum teleportation \cite{Togan2010,Hensen2015c}.

\subsection*{Initialization, Manipulation and Readout}
A combination of optical and microwave manipulation of quantum defects allows for full control of the spin. Quantum defects can be initialized into a well-defined ground state by various optical pumping schemes. For room-temperature operation, off-resonant excitation and spin-projection-dependent intersystem crossings (ISC) can be used to initialize the system \cite{Widmann2015}. This approach presents a simple route to initialization outside the low-temperature regime, but the fidelity is limited by the ISC rate contrast. At low temperatures, excitation of non-cycling fine-structure transitions may be used for high-fidelity ($>$99.7\%) initialization \cite{Robledo2011}. In this case, ISCs are not necessarily desirable, and may limit initialization fidelity. In the case of defects with non-integer spins $S>1/2$, it may be necessary to simultaneously drive spin transitions in order to create a well-defined initial state \cite{Nagy2019}.

Coherent control of the spin qubit is commonly achieved using microwave pulses, delivered by a wire loop antenna or stripline resonator, to directly drive transitions between spin states. The wide availability of ultra-stable commercial microwave sources allows the use of long dynamical-decoupling sequences to extend coherence times \cite{Abobeih2018a}.

Optical readout is one of the key advantages of optically-addressable quantum defects, enabling the qubit to function as a spin-photon interface. Long-distance quantum communication requires flying qubits, and contact-free readout is highly advantageous for many sensing applications. At low temperatures, excitation of a highly cycling transition provides a convenient route to high-fidelity readout, allowing for single-shot readout and quantum nondemolition measurements of the electron spin, with fidelities reaching 99.7\% \cite{Robledo2011,Sukachev2017a,Raha2019}. A key challenge is the extraction and detection of these photons from materials with high refractive indices. This problem can be overcome by micromachining of the solid-state host \cite{Hadden2010} and coupling to photonic structures \cite{Dibos2018,Hausmann2012,Gould2016,Barclay2011}.

Room-temperature readout typically relies on some spin-projection-dependent change in emitter brightness (\textit{e.g.} spin-dependent intersystem crossings competing with radiative relaxation). Readout contrasts in this scheme can vary significantly, from $\approx 30 \%$ in NV centers in diamond\cite{Steiner2010} to $< 1 \%$ for the nitrogen-vacancy in silicon carbide (NV$_\mathrm{SiC}$) defect \cite{Widmann2015}, depending on the relative radiative and non-radiative relaxation rates. Strategies have been developed to further improve readout fidelity in these systems, such as using spin-dependent ionization to increase contrast \cite{Shields2015a}, and repetitive readout schemes using nuclear spins as ancilla states \cite{Jiang2009}.

Electrical manipulation of point defects has been demonstrated in several systems. Spin-to-charge detection has been realized in NV centers in diamond \cite{Siyushev2019}, Si vacancies in SiC\cite{Niethammer2019}, and Er$^{3+}$ ions\cite{Yin2013}. Electric fields can also be used to modulate the optical properties of defects; Stark tuning of NV centers in diamonds can mitigate the effect of spectral diffusion \cite{Acosta2012}, while DC bias fields in SiC can be used to deplete nearby charge traps to stabilize the optical transition\cite{Anderson2019}. Another approach to qubit control focuses on using mechanical resonances to drive transitions between states\cite{Lee2017}; acoustically-driven strain has been used to manipulate spins in NV centers\cite{Macquarrie2013} in diamond and divacancy centers in SiC\cite{Tchebotareva2019}.

\subsection*{Material systems}

\subsubsection*{Currently studied systems}

While an exhaustive cataloging of optically-active defects in solid-state systems is beyond the scope of this review, several well-studied systems serve as exemplars of key concepts and design principles for various applications.

Diamond has proven to be one of the most successful host materials for quantum defects. The most-studied defect in diamond is the NV center, which has been widely explored as a sensor \cite{Steinert2010,Staudacher2013,Kucsko2013}, quantum register \cite{Dutt2007,Neumann2010,Bradley2019}, and quantum communication node \cite{Hensen2015c}. The room-temperature operation, biocompatibility, and photostability of the NV center make it an excellent sensor in a wide range of environments. However, the NV center faces several challenges in other quantum applications. It suffers from significant spectral diffusion of its optical transition\cite{Chu2014}, limited emission into its zero-phonon line (ZPL)\cite{Barclay2011}, and operates in a wavelength range that is not directly compatible with long-distance photon propagation, motivating broad searches for alternative defects. Nevertheless, the NV center remains best-in-class for spin coherence times, with $T_2$ times surpassing one second at low temperature\cite{Abobeih2018a}. This coherence time can be extended by mapping the electron spin coherence to a nuclear spin coherence of a nearby $^{13}$C, enabling repetitive readout protocols\cite{Jiang2009} and coherence times of seconds at room temperature\cite{Maurer2012}.

The negatively-charged silicon vacancy center (\sivm) has several desirable optical properties for quantum communication applications (limited spectral diffusion, high Debye-Waller factor\cite{Evans2016}), but requires cooling to millikelvin temperatures to overcome phonon-induced dephasing. At 4~K, the spin coherence is 35~ns \cite{Rogers2014a}, while at 10~mK, it can be extended by dynamical decoupling to 10~ms \cite{Sukachev2017a}. These properties have allowed for recent demonstrations of spin-photon entanglement using \sivm in nanophotonic structures\cite{Nguyen2019,Nguyen2019a}. Nearby nuclear spins can be used as ancilla qubits with coherence times of up to 0.2~s\cite{Nguyen2019}. The spin coherence can also be extended by increasing the energy splitting between spin-orbit states, for example by applying strain \cite{Sohn2018}. Other negatively charged group-IV vacancy defects have also been studied with the goal of achieving larger spin-orbit splittings \cite{Iwasaki2017,Trusheim2019,Siyushev2017,Trusheim2020}.

The neutral silicon vacancy (\sivo) has been shown to combine several of the desirable optical properties of \sivm with the long spin-coherence times of NV centers\cite{Rose2018a,Green2017a,Rose2018b}. This alternative charge state can be accessed by controlling the Fermi level of diamond\cite{Rose2018a}. It exhibits a spin coherence time of 255~ms with dynamical decoupling at 15 K, as well as near-transform-limited optical linewidths\cite{Rose2018a}. The wavelength of the \sivo~ZPL (946~nm) experiences significantly less attenuation in fibers, and is compatible with many frequency conversion schemes for telecom C-band wavelength operation\cite{Li2016}.

SiC shares several of the desirable material properties of diamond (large bandgap, low background magnetic noise, low spin-orbit coupling) and has also been shown to host many promising quantum defects. Coherence times of 1~ms and spin-dependent fluorescence have been demonstrated in the neutral divacancy and SiV$_{SiC}$ at cryogenic temperatures\cite{Christle2015,Nagy2019}. Room-temperature coherent manipulation of defects in SiC has also been demonstrated \cite{Widmann2015}. The near-IR operation (1300~nm) of the NV$_{SiC}$\cite{VonBardeleben2015} has low propagation loss in optical fibers, making it a promising candidate for long-distance quantum communication.

Rare-earth ions (REI) in solid-state hosts present a subtly different class of quantum defects. Transitions present in the free ion are perturbed by the crystal-field interaction with the host, but retain many of their ``atomic'' properties. Bulk REI samples have been studied extensively as potential ensemble quantum memories using photon echoes and other techniques\cite{Thiel2011}. Very recently, integration of nanophotonic structures with REI-doped crystals has enabled single-ion optical detection and manipulation\cite{Dibos2018,Zhong2018}, and key requirements for a quantum node have been demonstrated in single erbium ions, including single-shot readout \cite{Raha2019}. One significant challenge in the use of REIs as quantum resources is spin coherence time; this is often limited by other REI defects present in the crystal, or host nuclear spins \cite{McAuslan2012}. This motivates the search for new host materials, not only for REIs\cite{Phenicie2019a}, but for all types of quantum defects.

\subsubsection*{Prospects for new defects}
The primary function of the host material is to act as a structural matrix for the defect. A key requirement of a good host is that it introduces minimal additional noise into the system. Though there may be application-specific host requirements (\textit{e.g.} environmental compatibility considerations for sensors), the generally-desirable properties of the host may be summarized as:

\begin{enumerate}
	\item Bandgap sufficient to support the defect optical transition.
	\item Minimal magnetic noise background: low paramagnetic defect concentration, low nuclear spin background. The latter consideration limits the possible materials to those entirely comprised of elements with only nuclear-spin-zero isotopes.
	\item High Debye temperature and/or low spin-orbit coupling to minimize relaxation and dephasing.
\end{enumerate}

The optimal properties of a defect are typically application-specific (\textit{e.g.} optimal wavelength operation, importance of room-temperature operation), though long spin coherence times and stable optical transitions are generally desirable in all applications.

\subsection*{Applications}
\subsubsection*{Quantum Sensing} Optically addressable quantum defects have been utilized extensively in quantum sensing; key advantages are the contact-free mode of operation (requiring only an optically-transparent environment) and the high spatial resolution offered by their point-like nature. Additionally, room-temperature coherence and straightforward integration into nanoparticles make these qubits especially attractive as \textit{in situ} sensors.

One natural application of these defects is as magnetometers. DC vector magnetometry has been demonstrated with sensitivities of 500~nT/$\sqrt{\mathrm{Hz}}$\cite{Liu2019} using single defects and 50~pT/$\sqrt{\mathrm{Hz}}$\cite{Schloss2018} in ensemble samples. Color centers have also shown great utility sensing AC magnetic fields, achieving sensitivities of 4.3~nT/$\sqrt{\mathrm{Hz}}$\cite{Balasubramanian2009}, limited by the coherence time of the defect. Further improvements in sensitivity have been demonstrated using quantum logic\cite{Jiang2009,Lovchinsky2016} and spin-to-charge readout\cite{Shields2015a}. Sensitivity to small fields is not the only important metric for AC magnetometers, frequency resolution and precision are also key parameters. Recent pulse sequences have been developed to enhance the frequency resolution beyond limits imposed by NV spin decoherence, and have achieved $<100~\mu$Hz frequency resolution and $<\mu$T precision\cite{Boss2017,Schmitt2017a}, offering a route to single-molecule chemical-shift nuclear magnetic resonance experiments\cite{Aslam2017}. Combining NV-containing nanodiamonds with scanning probe microscopes enables nanometer scale imaging\cite{Laraoui2015}. An alternative approach is to construct atomic-force microscope cantilevers directly from diamonds containing color centers\cite{Pelliccione2016,Zhou2017}.

Beyond magnetic field sensing, the high thermal conductivities of diamond and SiC have led to their use as nanoscale temperature sensors. Both microwave-assisted and all-optical thermometry schemes with NV centers have been demonstrated\cite{Kucsko2013,Fukami2019}, while optical thermometry schemes with the silicon-vacancy defects in diamond\cite{Nguyen2018a} and SiC\cite{Anisimov2016} have been reported with sensitivities down to 100~mK$/\sqrt{\mathrm{Hz}}$. These schemes rely on subtle changes in the host lattice with temperature, and so do not require any additional labelling. Note that these defects all operate in distinct wavelength ranges; optical multiplexing could be used to increase the sensitivity further.

A key challenge remaining in the use of quantum defects as nanoscale sensors is gaining control over the surface of the sensor. Both sensitivity and resolution require that the defect must be close to the sensing target, which necessitates being near the host surface. This requirement brings significant material engineering challenges; surface-related defects such as charge traps and dangling bonds can lead to Fermi level pinning and magnetic noise, degrading charge stability and coherence time of near-surface quantum defects. Many groups have observed that NV coherence degrades as NV centers are brought closer to the surface \cite{Myers2014,FavarodeOliveira2017,Sangtawesin2018a1}. Recent work has allowed for the realization of NV centers with coherence times exceeding 100 $\mu$s within nanometers of the surface by careful control over the diamond surface \cite{FavarodeOliveira2017,Sangtawesin2018a1}.

\subsubsection*{Quantum Simulation}
The straightforward initialization and control of quantum defects suggest they may find a natural home in quantum simulation applications. However, the probabilistic nature of formation of these defects makes the construction of a target Hamiltonian extremely challenging. An additional challenge is the length scale of spin-spin interactions (several nanometers) and the minimum defect separation imposed by the optical diffraction limit (hundreds of nanometers), though this challenge may be overcome with spectrally-distinct emitters \cite{Raha2019}.

Instead, several proposals have focused on utilizing nearby spins to serve as a simulator, with the optically active defect serving as a route to both manipulating and reporting on these proximal spins. Proposed systems include using dipolar interactions between nitrogen spins in the bulk as buses\cite{Yao2012}, utilizing nearby spins at a diamond surface to probe many-body localization\cite{Serbyn2014}, and patterning nuclear spins on a surface to realize effects such as frustrated magnetism\cite{Cai2013}. A key technical achievement to this end has been mapping out the locations of nearby $^{13}$C spins in diamond, and using this information to manipulate individual nuclear spins \cite{Bradley2019,Abobeih2019}.

Quantum defects find broader applicability when instead of simulating arbitrary target systems, they are instead used to probe new physics in disordered systems. One example of this approach is the experimental realization of time crystals using \nvm{} centers in diamond\cite{Choi2017}.

\subsubsection*{Quantum Computation}
Many of the Di Vincenzo criteria have been robustly met in quantum defect systems; optical pumping provides high-fidelity intialization\cite{Rogers2014a,Raha2019,Sukachev2017a}, long spin coherence times have been demonstrated in many systems\cite{Christle2015,Sukachev2017a,Rose2018a,Abobeih2019}, spin-selective transitions provide a straightforward readout channel, and two-qubit gates have been demonstrated using nearby nuclear spins\cite{Zu2014}. As with simulation applications, however, scalability of these systems is hampered by the probabilistic formation of defects and the positions of the nuclear spins surrounding them. 

Though these defects may not be suited to large-scale quantum computation, the use of smaller quantum registers\cite{Dutt2007} may be advantageous in modular quantum computing, enhanced sensing, or high-fidelity quantum communication. Key advances to this end are the development of protocols enabling the manipulation of nearby nuclear spins via hyperfine interactions with NV centers, enabling registers with tens of qubits\cite{Bradley2019,Abobeih2019}.

\subsubsection*{Quantum Communication} The ability to serve as a spin-photon interface makes quantum defects uniquely well-suited for quantum communication applications. The spin degree of freedom can serve as a local quantum memory, while the photon takes the role of a flying qubit; remote entanglement over the kilometer scale has already been demonstrated with defects in diamond\cite{Hensen2015c}.

A common scheme used to generate entanglement between two defects uses the interference of indistinguishable photons on a beamsplitter. An optimal defect for this application would have a high rate of generation of indistinguishable photons, and these photons would experience minimal loss over the distance they propagate. Though NV centers in diamond have been the main platform used to demonstrate these entanglement experiments \cite{Bernien2013,Hensen2015c}, the large fraction of the emission ($>0.97$) into the phonon sideband\cite{Barclay2011}, spectral diffusion of the optical transitions, and the high attenuation of 637-nm light through fibers on the kilometer scale precludes the ``as-is'' use of NV centers in a large quantum network. Efforts to address this include nanophotonic structures to enhance the ZPL emission\cite{Gould2016,Barclay2011}, and frequency conversion to telecom wavelengths\cite{Dreau2018}.

Other quantum defects have been proposed as alternatives to NV centers. \sivm and \sivo show reduced spectral diffusion because of their inversion symmetric structures\cite{Sipahigil2014,Rose2018a}, allowing for higher entanglement generation rates. Silicon vacancies in SiC have also shown limited spectral diffusion\cite{Nagy2019}; here, it is the similarity between permanent electric dipole moments in the ground and excited states that minimizes spectral diffusion. Several key technical accomplishments for quantum communication have been demonstrated with the \sivm defect. Cavity-mediated interactions between different defects have been demonstrated \cite{Sipahigil2016,Evans2018}, as have spin-photon interfaces and the formation of local quantum registers\cite{Nguyen2019,Nguyen2019a}. Scalable fabrication of the nanophotonic cavities used in these demonstrations is an important step in the development of \sivm for quantum communication \cite{Wan2019}. However, the limited spin coherence time and millikelvin operation temperature of \sivm is a significant challenge for building a large network, motivating the development of the \sivo defect. Though this system has better coherence time at 4~K \cite{Rose2018a}, thus far, spin-dependent fluorescence has not been observed in \sivo, which is a requirement for establishing a spin-photon interface. 

For long-range quantum communication, operation in the 1550-nm wavelength range is essential. \erb ions are one of the few systems that have optical transitions in this spectral region. Detection of single ions has been historically limited by the slow emission rate ($\approx 100$~Hz); however, integrated nanophotonic cavities have demonstrated significant Purcell-enhanced emission rates can be achieved\cite{Dibos2018,Raha2019}, presenting a route towards long-range, single-ion, \erb-based quantum repeaters.

\section*{Outlook}
In this review we have highlighted the variety of quantum systems and applications offered by semiconductor qubits. With high fidelity initialization and readout methods, long coherence times with fast gate operation, inter-qubit coupling and coupling to photons, many of these systems fulfill criteria essential for quantum computation proposals. We envision that a diverse ecosystem of qubits will allow for many different quantum applications, with inherent advantages and trade-offs for each qubit. The semiconductor community is very far from having `picked a winner'. As this review has highlighted, `winner' may never even need to be picked; rather, the diverse and eclectic nature of semiconductor system is likely to remain a defining feature of this platform. This situation is rather different from e.g. superconducting qubits, where the community has mostly converged towards a rather narrow portfolio of physical systems, optimized for quantum information processing.

\begin{figure*}
	\includegraphics[width=\textwidth]{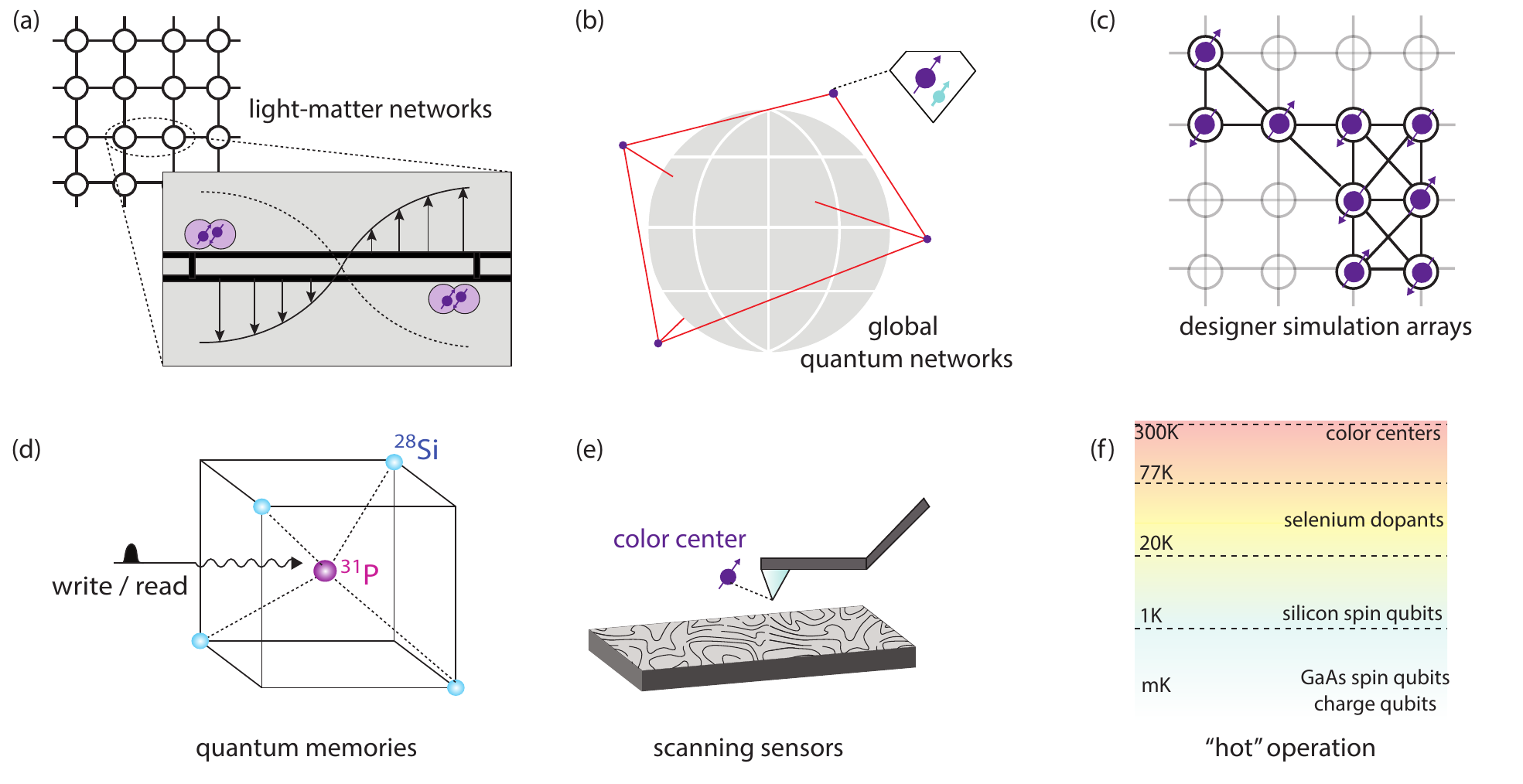}
	\caption{
		\textbf{Future outlook for semiconductor qubits}. 
		The applications for semiconductor qubits are extremely diverse, and we show but a few here. (a) The demonstration of strong coupling between microwave photons and charge- or spin-qubits in gate-controlled systems underpins the field of hybrid systems that blend light and matter, or superconductor-semiconductor networks. The inset shows a network with local nodes consisting of quantum-dot-based spin qubits coupled via a superconducting resonator. (b) Color centers have made headway towards quantum networks, with the ability to entangle distant qubits across hundreds of kilometers. Other uses include entanglement beacons for cryptography and secure global communications. (c) Quantum-dot based arrays of spin qubits have been used for proof-of-principle simulations of condensed-matter systems such as the Hubbard model and Nagaoka ferromagnetism. 2D arrays and arbitrary connections could enable versatile, reconfigurable simulations. (d) Dopants in silicon have shown extraordinary coherence times, and are among of the best candidates for addressable quantum memories and dense qubit arrays. (e) Color centers fabricated on the tip of a scanning probe can be used for extremely sensitive magnetometry across surfaces. (f) An attractive proposition is the operation of qubits at temperatures reachable without dilution refrigerators or even at room temperature. While most gate-controlled spin and charge qubits require millikelvin temperatures, spins in silicon have been operated as qubits above 1~K, while selenium dopants in silicon are predicted to be operable above 77~K. Color centers, indeed, can be operated at room temperature.
	}
	\label{fig5}
\end{figure*}

Therefore, Figure~\ref{fig5} illustrates a range of few applications that are likely to be pursued using the broad palette of semiconductor qubit encodings. Charge-photon and spin-photon interfaces with strong coupling have been recently demonstrated (and reviewed here\cite{Burkard2020}), and advances in resonator engineering may lead to light-matter networks (panel a), entangling distant qubits for a surface code and other qubit-qubit connectivities.
A quantum internet (panel b) based on secure quantum cryptography and entanglement beacons may become possible when semiconductor qubits are made optically active. Panel (c) shows a powerful application of gate-controlled spin qubits in quantum dots, where specific two-dimensional arrays implement Hamiltonians desired for quantum simulations. 
Dopant atom with long-lived spin states can be used to store information as quantum memories (panel d), including in ensembles coupled to microwave cavities, or in hybrid donor-dot systems. Quantum sensing applications are already widely researched, with nanomechanical piezoelectric actuators used to create scanning quantum probes (panel e) for magnetism and novel spin textures. Lastly, quantum systems operable above millikelvin-scale temperatures would go a long way toward solving the problem of wiring and cooling large numbers of physical qubits; panel (f) shows a few semiconductor qubits that have already be operated above dilution-refrigerator temperatures.

Several challenges remain for semiconductor quantum circuits. Probably the most significant one is the establishment of scalable and reproducible fabrication processes. This task is rendered onerous by the extremely small physical dimensions of semiconductor qubits, but the future prospects are brightened by the potential for integration with industry-based manufacturing methods. 
Calibrating, operating and stabilising large arrays, or simultaneously performing and validating quantum operations over multiple qubits are among the current challenges facing quantum computing research groups. The broad issues are superficially similar to those being attacked by superconducting circuits and ion traps, but the small size of semiconductor qubits poses unique problems with cross-talk and placement of classical readout devices. In return, the extreme density afforded by semiconductor systems lends credibility to the prospect of integrating hundreds of millions of qubits on a chip, as necessary for the most useful quantum algorithms integrated with quantum error correction. 
For quantum sensing, the next frontier is developing methods to operate in three dimensions and under an ever wider range of environmental conditions.

In conclusion, the diversity and flexibility afforded by semiconductor materials and encodings will continue to encourage the community to work in many complementary directions. An increasing engagement with the semiconductor industry will enable exceptional levels of qubit density and device reliability; this will not only enable the production of useful and manufacturable quantum devices, but also continue to expand the scope for studying fundamental science in engineered quantum systems of unprecedented complexity.

\section*{Acknowledgements}
A.C and F.K. acknowledge support from the European Union's Horizon 2020 research and innovation programme under grant agreement MOS-quito (No. 688539). A.C. acknowledges support from the EPSRC Doctoral Prize Fellowship. 
S.D.F. acknowledges support from the European Union, through the Horizon 2020 research and innovation program (Grant Agreement No. 810504), and from the Agence Nationale de la Recherche, through the CMOSQSPIN project (ANR-17-CE24-0009).
A.M. acknowledges funding from the Australian Research Council (Projects CE170100012 and DP180100969), the U.S. Army Research Office (Grant no. W911NF-17-1-0200) and Australian Department of Industry, Innovation and Science (Grant no. AUSMURI00002).
F.K. acknowledges support from the Independent Research Fund Denmark. 
N.P.d.L. acknowledges support from the NSF under the EFRI ACQUIRE program (grant 1640959) and the CAREER program (Grant No. DMR-1752047), the Air Force Office of Scientific Research (award numbers FA9550-17-0158 and FA9550-18-1-0334), the Eric and Wendy Schmidt Transformative Technology Fund, and the Princeton Catalysis Initiative.



\end{document}